\documentclass[10pt,aps,prb,twocolumn]{revtex4-1}
\usepackage{graphicx}
\usepackage{amssymb}
\usepackage{amsmath}
\usepackage{appendix}
\usepackage{comment}

\newcommand{\eq}[1]{\begin{equation}#1\end{equation}}

\usepackage{mdframed}

\begin{document}

\title{Perspective: (Beyond) spin transport in insulators}

\author{Yaroslav Tserkovnyak}
\affiliation{Department of Physics and Astronomy, University of California, Los Angeles, California 90095, USA}

\begin{abstract}
Insulating materials with dynamical spin degrees of freedom have recently emerged as viable conduits for spin flows. Transport phenomena harbored therein are, however, turning out to be much richer than initially envisioned. In particular, the topological properties of the collective \textit{order-parameter textures} can give rise to \textit{conservation laws} that are not based on any specific symmetries. The emergent continuity relations are thus robust against structural imperfections and anisotropies, which would be detrimental to the conventional spin currents (that rely on approximate spin-rotational symmetries). The underlying fluxes thus supersede the notion of spin flow in insulators, setting the stage for nonequilibrium phenomena termed \textit{topological hydrodynamics.} Here, we outline our current understanding of the essential ingredients, based on the \textit{energetics} of the electrically-controlled injection of topological flows through interfaces, along with a reciprocal signal generation by the outflow of the conserved quantity. We will focus on two examples for the latter: winding dynamics in one-dimensional systems, which supplants \textit{spin superfluidity} of axially-symmetric easy-plane magnets, and \textit{skyrmion} dynamics in two-dimensional Heisenberg-type magnets. These examples will illustrate the essential common aspects of topological flows and hint on generic strategies for their generation and detection in spintronic systems. Generalizations to other dimensions and types of order-parameter spaces will also be briefly discussed.
\end{abstract}

\maketitle

\section{Introduction}

Understanding electricity, which concerns phenomena deriving from the motion of electric charge, has been a cornerstone of solid-state physics. Studying and quantifying such motion, e.g., through the measurements of electrical conductivity, provided fundamental probes of materials that lead to some of the central discoveries of the 20th-century physics, such as superconductivity and quantum Hall effect. Being primarily carried by electrons, electric charge flows can be used to differentiate between some of the basic electronic states of crystals, such as metals, insulators, and semiconductors. Generally, whenever electronic charge correlations bear some key signatures of the underlying phase or state of a material, we can expect the electrical conductivity to offer a valuable probe thereof. Conversely, a material known to have some striking electrical response can be tailored for electronic applications.

A broad range of complex materials, however, have their key dynamic properties rooted in different physics. In particular, magnetic materials may have essentially no electrical response, up to high frequencies (determined by the gap for charge excitations), while having their prevalent low-frequency fluctuations governed by the spin degrees of freedom. This concerns, more generally, systems with strong spin correlations and/or frustration, where the low-energy properties are either dominated or, at least, strongly affected by the correlated spin dynamics.

Spintronics has recently emerged as a field that exploits these spin degrees of freedom to either study the underlying materials and heterostructures or employ the associated functionality in novel devices and computing architectures.\cite{wolfSCI01,*zuticRMP04,*sinovaNATM12,*baltzRMP18} One feature that distinguishes spintronics from other spin-based disciplines, such as various spin-resonance and scattering spectroscopies, is a focus on transport regimes, where the net spin angular momentum in the system is conserved. In this case, supported by the reasoning that is similar to that underlying Kirchhoff's circuit laws for charge flows in electrical circuits, one can construct spin-flow-based principles for spin dynamics.\cite{tserkovRMP05} Interfaces or junctions in a spin-active heterostructure would then serve as nodes that transmit spin flows.\cite{brataasPRL00} The spin flow over a certain region (e.g., an interface between two materials or a section of a single material), which serves as a basic building block for the circuit perspective, can be driven by an effective spin bias. Thermodynamically, the latter can be understood as a drop in the spin (chemical) potential, which is locally conjugate to the spin density. The spin conservation would dictate a homogeneity of the spin potential in equilibrium.

While a finite spin flow across a heterointerface may have to be transmuted between physically disparate entities, such as electron-hole pairs on one side and magnons on the other,\cite{bauerPHYS11,benderPRL12} it can still be conserved. Such conservation, along a specific axis, relies in general on the corresponding spin-rotational symmetry, which must be satisfied in both materials as well as at the interface itself. In practice, this is of course always an approximation, which might explain why the basic notion of spin transport\cite{dyakonovJETPL71} was not widely accepted for a long time. One important issue here is that spin signals carried by decaying quasiparticles are exponentially suppressed beyond the associated spin-diffusion length.\cite{zuticRMP04}

In this Perspective, I will start by recapping some recent developments in our understanding of spin flows through magnetic insulators. We will initially suppose that the spin bias is produced by a nonequilibrium electron spin accumulation, which can be controlled electrically.\cite{hirschPRL99,*hoffmannIEEEM13,*sinovaRMP15,brataasPRP06} It turns out, however, that an ordinary spin flow is not the only transport process that can be triggered by such spin biases. Thinking more broadly about the coherent (magnetic) order-parameter dynamics, which can be controlled and detected electrically, will bring us to the notion of the conserved topological flows. An idealized concept of spin superfluidity\cite{halperinPR69,*soninJETP78,*soninAP10} is perhaps the simplest example thereof, which will be relied heavily on for pedagogical purposes. We will discuss how the interplay of current-induced work, topology, and coherent spin dynamics can conspire to yield robust long-distance and low-dissipation information flows through magnetic insulators.

\section{Background}

\subsection{Spin-flow nodes and circuitry}

\begin{figure}[!b]
\includegraphics[width=0.7\linewidth]{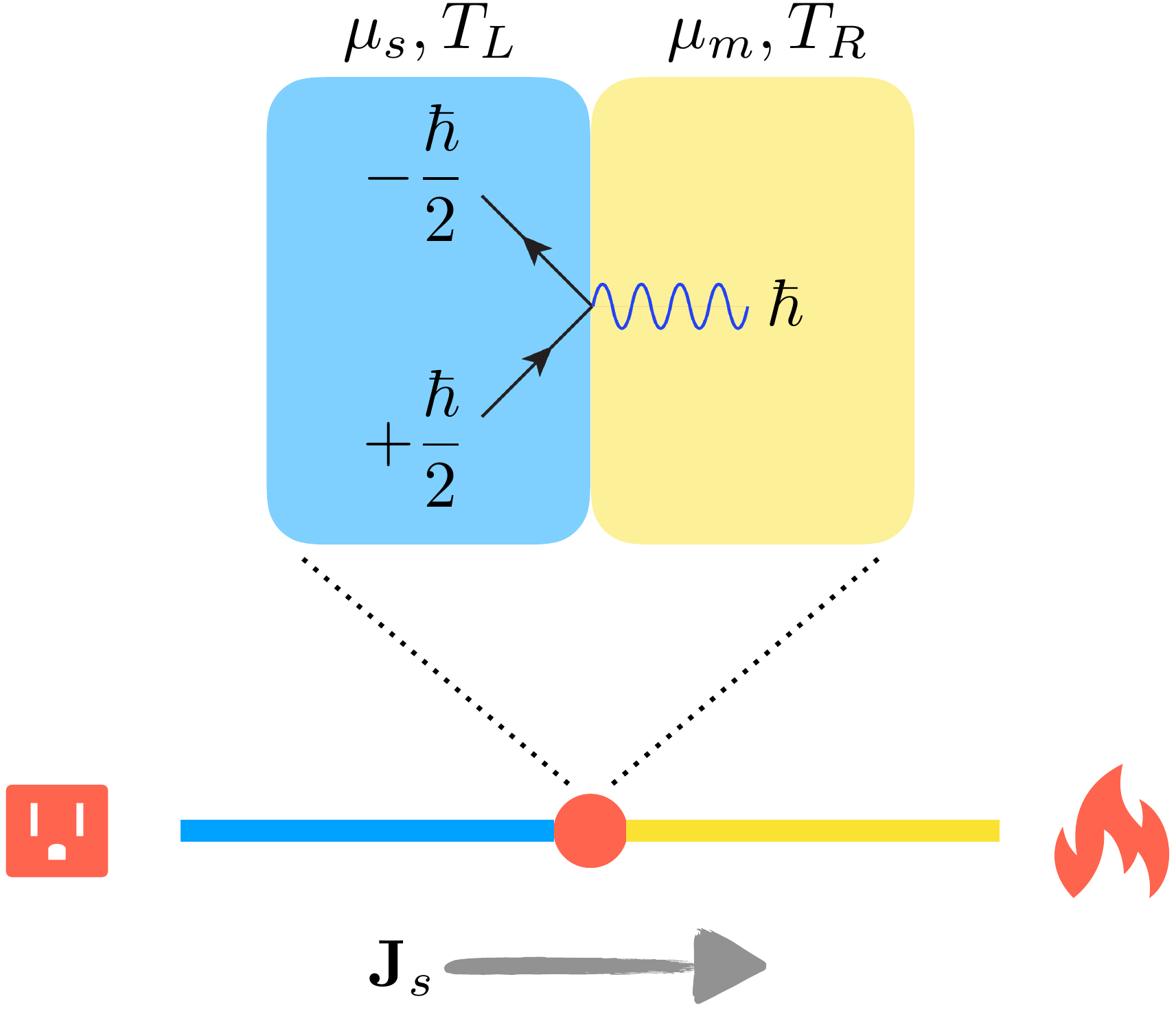}
\caption{A schematic of an elementary spin-transport node between a nonmagnetic metal (left) and a magnetic insulator (right). Electrons can flip their spin at the interface, while transmitting (or absorbing) a magnon. The spin current $J_s$ is driven by the thermodynamic bias of $(\mu_s-\mu_m)$, in spin sector, and $(T_L-T_R)$, in heat sector. Such a bias across the interfacial node can be established in response to thermoelectric controls of a larger circuit, in a self-consistent steady state.\cite{bauerNATM12,benderPRB15}}
\label{sch}
\end{figure}

In a simple illustration of spin flows in solid-state heterostructures, consider a junction between a nonmagnetic metal and a magnetic insulator, as depicted in Fig.~\ref{sch}. This junction can be viewed as a node in a larger circuit, which could be ultimately driven by a combination of electrical and thermal means (through, e.g., the so-called spin Hall\cite{dyakonovJETPL71,hirschPRL99} and spin Seebeck\cite{uchidaNATM10,*bauerNATM12} effects, respectively). In a nonequilibrium steady state, we can have a situation, in which the itinerant electrons in the metal obey the Fermi-Dirac statistics with the spin-dependent distribution function
\eq{
n_{\rm FD}(\epsilon)\stackrel{\uparrow/\downarrow}{=}\frac{1}{e^{\beta_L(\epsilon\mp\mu_s/2)}+1}\,,
\label{nFD}}
while the magnons follow the Bose-Einstein distribution
\eq{
n_{\rm BE}(\epsilon)=\frac{1}{e^{\beta_R(\epsilon-\mu_m)}-1}\,.
}
$\beta\equiv1/k_BT$ stands for the inverse temperature, on each side, $\mu_s$ is the spin potential (also known in the literature as the spin accumulation\cite{brataasPRL00,tserkovRMP05,brataasPRP06}) in the metal, while $\mu_m$ is the spin potential (which corresponds simply to the bosonic chemical potential\cite{benderPRL12}) in the magnet. Orienting the spin quantization axis here along a symmetry axis in spin space (which, in the case of a collinear spin order, must be along the order parameter), the spin flow is continuous across the interface. In linear response, it should generally obey the following phenomenology:
\eq{
J_s=G(\mu_s-\mu_m)+S(T_L-T_R)\,,
\label{Jth}}
in close analogy with thermoelectricity.\cite{mahanBOOK00} $G$ here is the interfacial spin conductance and $S$ is the spin Seebeck coefficient. In thermodynamic equilibrium, $\mu_s=\mu_m$ and $T_L=T_R$, so that $J_s=0$. Microscopically, the values of $G$ and $S$ depend on the strength of the (Heisenberg) spin exchange at the interface, between the itinerant electron spins on the left and localized magnetic moments on the right.\cite{xiaoPRB10,adachiPRB11,benderPRL12,rezendePRB14,benderPRB15} These parameters, furthermore, depend on the ambient temperature, typically increasing with temperature, due to the bosonic statistics of magnons.

\subsection{Energetics of the coherent spin transfer}

Let us now look into the process of spin injection at an interface between a normal metal and a dynamic magnet. At sufficiently low temperatures, we can neglect thermal spin excitations, like those underlying Eq.~\eqref{Jth}, and instead focus on the coherent spin dynamics as well as the spin transport driven by a (vectorial) spin bias $\boldsymbol{\mu}_s$ in the normal metal.\cite{tserkovRMP05} Its absolute value is $|\boldsymbol{\mu}_s|=\mu_s$ and the direction is determined by the spin-quantization axis for which the electron occupation follows Eq.~\eqref{nFD}.

As a starting point, consider a simple collinear ordering in the magnet, whose dynamic state is described by a directional order parameter $\mathbf{l}(t)$, s.t. $|\mathbf{l}(t)|\equiv1$. Writing the (vectorial) spin current $\mathbf{J}_s$ across the interface in terms of $\boldsymbol{\mu}_s$ and a slowly-varying $\mathbf{l}(t)$ then gives\cite{tserkovRMP05}
\eq{
\mathbf{J}_s=\frac{g}{2\pi}\mathbf{l}\times\left(\boldsymbol{\mu}_s\times\mathbf{l}-\hbar\dot{\mathbf{l}}\right)\,.
\label{Js}}
$\mathbf{l}$ here can physically stand for the magnetic order in a ferromagnet or the N{\'e}el order in an antiferromagnet.\cite{takeiPRL14,*takeiPRB14} The interfacial coefficient $g$ is known as the spin-mixing conductance.\cite{brataasPRL00,tserkovRMP05,brataasPRP06} The expression \eqref{Js} is isotropic in spin space, obeys Onsager reciprocity\cite{onsagerPR31p1,*onsagerPR31p2} (when viewed as relating the spin flow into the normal metal with the order-parameter dynamics in the magnet\cite{takeiPRL14}), and vanishes when the frequency of rotation matches the spin bias (which is easily understood in the rotating frame of reference\cite{tserkovRMP05}). This expression, furthermore, breaks the (macroscopic) time invariance, as $\mathbf{J}_s\to\mathbf{J}_s$, $\mathbf{l}\to-\mathbf{l}$, and $\boldsymbol{\mu}_s\to-\boldsymbol{\mu}_s$, under time reversal. This underlines its dissipative character, which we can exploit in order to pump energy into the magnetic dynamics.

Spin transfer \eqref{Js} across the interface signifies a torque, $\mathbf{J}_s\to\boldsymbol{\tau}$, when viewed from the point of view of the magnetic dynamics, which translates into work
\eq{
\dot{W}\equiv\boldsymbol{\tau}\cdot\mathbf{l}\times\dot{\mathbf{l}}=\frac{g}{2\pi}\left(\boldsymbol{\mu}_s\times\mathbf{l}-\hbar\dot{\mathbf{l}}\right)\cdot\dot{\mathbf{l}}
\label{W}}
on the magnetic order, per unit time. The second term, $\propto-(\dot{\mathbf{l}})^2$, on the right-hand side contributes to the generic Gilbert damping\cite{gilbertIEEEM04} of the magnetic dynamics, while the first term, which is sometimes referred to as the antidamping torque,\cite{ralphJMMM08} may effectively reverse the sign of the natural damping, leading to a dynamic instability. We can understand Eq.~\eqref{W} from the Hamilton equations of motion for the order-parameter $\mathbf{l}$ dynamics. To this end, we modify the rate of change of the conjugate momentum $\boldsymbol{\pi}_\mathbf{l}$ as
\eq{
\dot{\boldsymbol{\pi}}_\mathbf{l}=-\frac{\partial H}{\partial\mathbf{l}}+\boldsymbol{\tau}\times\mathbf{l}\,,
}
in the presence of an interfacial torque $\boldsymbol{\tau}$, where $H(\mathbf{l},\boldsymbol{\pi}_\mathbf{l})$ is the Hamilton function. The reason for this is that $\boldsymbol{\pi}_\mathbf{l}=\boldsymbol{\rho}_s\times\mathbf{l}$, with the spin density $\boldsymbol{\rho}_s$ being the generator of rotations.\cite{andreevSPU80} Its dynamics are modified by the spin torque as $\dot{\boldsymbol{\rho}}_\mathbf{l}\to\dot{\boldsymbol{\rho}}_\mathbf{l}+\boldsymbol{\tau}$. The work production \eqref{W} by the torque is then finally obtained as $\dot{H}=\dot{\boldsymbol{\pi}}_\mathbf{l}\cdot\partial H/\partial\boldsymbol{\pi}_\mathbf{l}+\dot{\mathbf{l}}\cdot\partial H/\partial\mathbf{l}=\boldsymbol{\tau}\cdot\mathbf{l}\times\dot{\mathbf{l}}$, invoking also the other Hamilton equation: $\dot{\mathbf{l}}=\partial H/\partial\boldsymbol{\pi}_\mathbf{l}$.

\begin{figure}[!htb]
\includegraphics[width=0.8\linewidth]{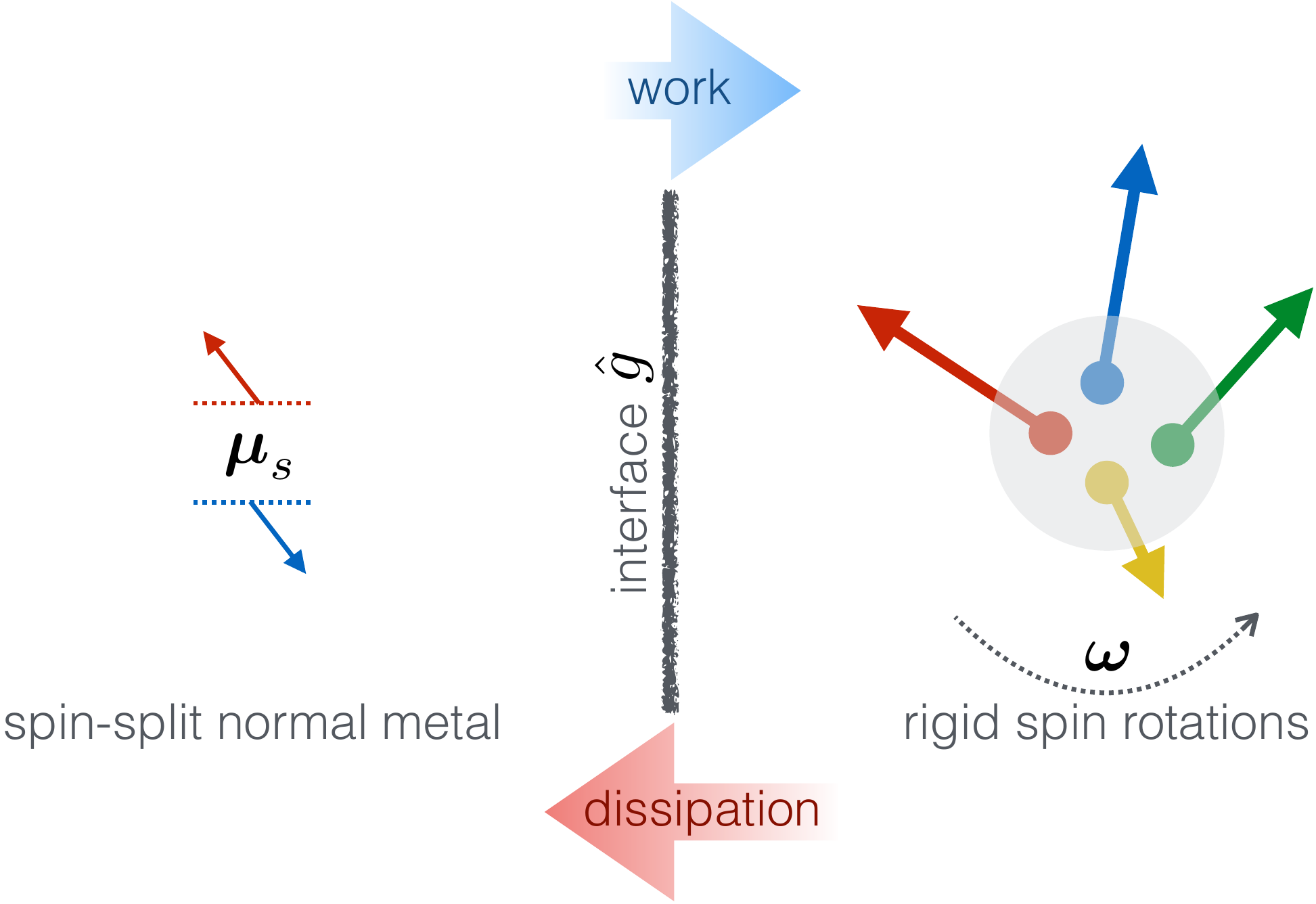}
\caption{A schematic of a noncollinear spin system (right) in contact with a metallic spin reservoir (left). The nonequilibrium spin state of the metal is parametrized by the (vectorial) spin accumulation $\boldsymbol{\mu}_s$. The magnet, whose spin arrangement is determined by some isotropic exchange Hamiltonian, is described, near the interface, by uniform (and essentially rigid) rotations of all spins. Its instantaneous nonequilibrium state is thus characterized by the (vectorial) frequency of SO(3) rotation $\boldsymbol{\omega}$. The $3\times3$ matrix $\hat{g}$ generalizes the concept of the spin-mixing conductance $g$ pertinent to the collinear case. The figure is adapted from Ref.~\onlinecite{tserkovPRB17}.}
\label{gen}
\end{figure}

More generally, for a noncollinear spin order that can be parametrized by an SO(3) rotation matrix $\hat{R}(t)$, the appropriate torques in the equation of motion can be derived from the following Rayleigh dissipation function:\cite{tserkovPRB17}
\eq{
R=\frac{1}{2h}(\boldsymbol{\mu}_s-\hbar\boldsymbol{\omega})\hat{g}(\boldsymbol{\mu}_s-\hbar\boldsymbol{\omega})\,.
\label{R}}
which corresponds to the (half of the) net dissipation in the combined nonequilibrium system (i.e., the magnet plus the adjacent metal). Here, $h\equiv2\pi\hbar$ and $\boldsymbol{\omega}\equiv i\mbox{Tr}[\hat{R}^T\hat{\boldsymbol{L}}\partial_t\hat{R}]/2$ is the (vectorial) angular velocity of the spin dynamics, defined in terms of the vector $\hat{\boldsymbol{L}}$ of SO(3) generators: $\{\hat{L}_\alpha\}_{\beta\gamma}\equiv-i\epsilon_{\alpha\beta\gamma}$, the Levi-Civita symbol. $\hat{g}$ is a symmetric real-valued $3\times3$ matrix, whose diagonalization defines three principal axes along with the associated (nonnegative) spin conductances, which generalize the scalar (spin-mixing) conductance $g$ discussed above. This treatment may be applied, e.g., to noncollinear antiferromagnets and spin glasses with an (effective) SU(2) symmetry.\cite{tserkovPRB17,ochoaPRB18} In the simplest case of an isotropic spin glass, $\hat{g}\propto\hat{1}$. Figure \ref{gen} shows a schematic of the nonequilibrium system at hand. The Rayleigh dissipation function \eqref{R} encodes the information about the dissipation of the magnetic dynamics into the normal-metal reservoir as well as the reciprocal work done by a nonequilibrium spin accumulation $\boldsymbol{\mu}_s$ applied to it.\cite{tserkovPRB17}

In closing this section, we would like to recall that a straightforward way to establish an effective spin accumulation $\boldsymbol{\mu}_s$ at a boundary of a generic conductor is by using the spin Hall effect.\cite{dyakonovJETPL71,hirschPRL99} Namely, on general symmetry grounds, we may write
\eq{
\boldsymbol{\mu}_s=\vartheta_{\rm sH}\,\mathbf{z}\times\mathbf{j}\,,
\label{sH}}
where $\mathbf{z}$ is the normal to the interface and $\mathbf{j}$ is the (tangential) electric current density. $\vartheta_{\rm sH}$ is a material-dependent parameter that depends on the strength of spin-orbit interactions near the interface, vanishing in the absence thereof. Some heavy metals and, particularly, the so-called topological-insulator materials are known to engender a sizable $\vartheta_{\rm sH}$.\cite{hellmanRMP17}

In the presence of a proximal magnetic material, which modifies the spin-related boundary condition according to, e.g., Eq.~\eqref{Js}, the spin accumulation $\boldsymbol{\mu}_s$ generally needs to be calculated self-consistently, together with solving the magnetic equations of motion.\cite{tserkovRMP05} In certain special cases, however, particularly in the limit of very fast spin relaxation in the metal, the latter may be treated as a good spin reservoir that is not significantly affected by the spin flow in and out of the adjacent magnet.

\section{Towards topological field flows}

\subsection{Spin flow through an arbitrary insulator}

\begin{figure}[!htb]
\includegraphics[width=0.99\linewidth]{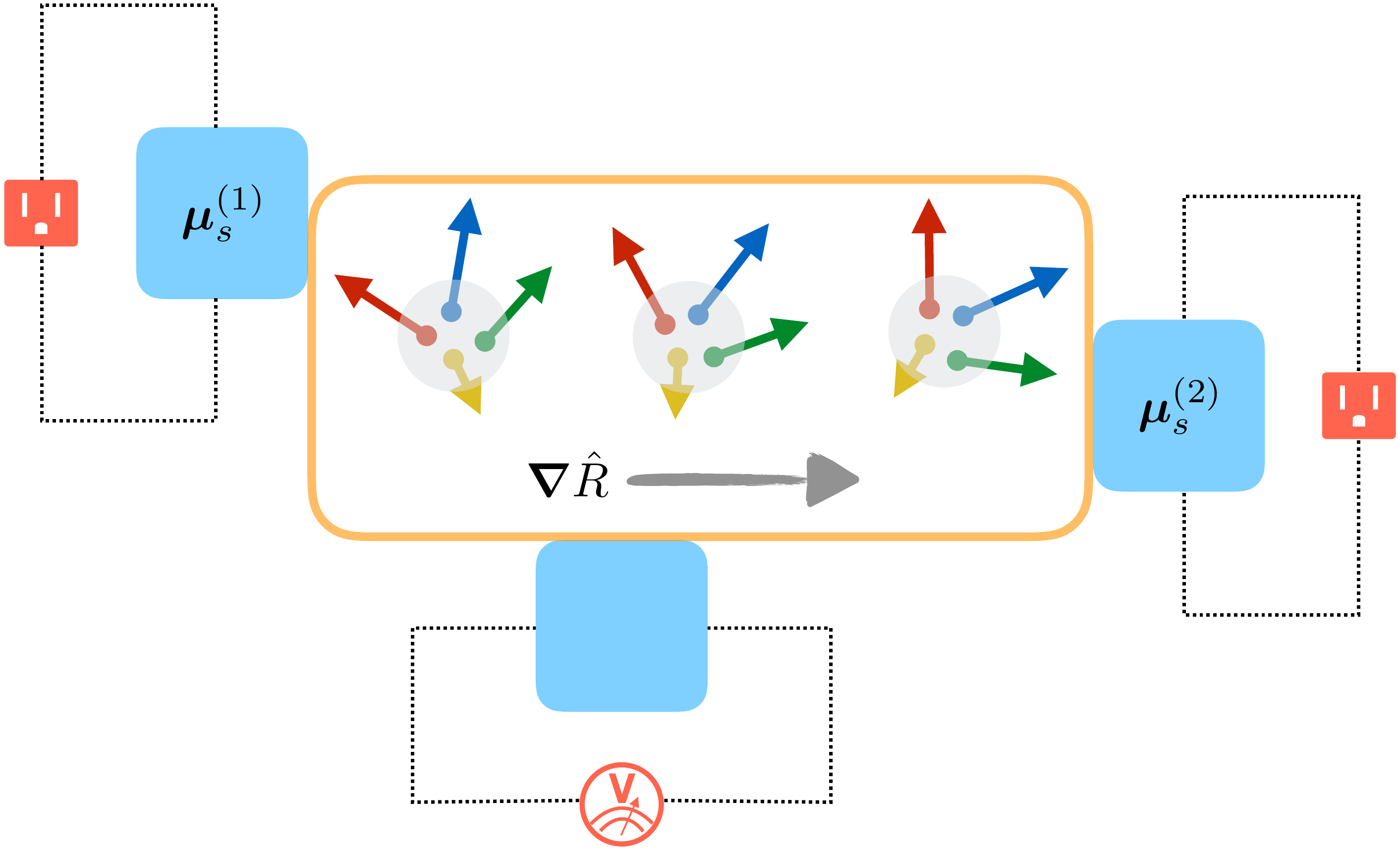}
\caption{A general spin circuit, in which the input terminals labelled by $i=1,2,\dots$ establish local spin biases $\boldsymbol{\mu}^{(i)}_s$, which drive spin dynamics in the material or system of interest. The readout terminal (bottom) performs a measurement of the resultant dynamic steady state via reciprocal means. For example, if the input is accomplished by the spin Hall effect, by applying electrical currents that induce spin accumulations \eqref{sH}, the output can be achieved by measuring the inverse spin Hall voltage.\cite{hirschPRL99} The instantaneous state of the driven system can be described, e.g., by the position-dependent rotation matrix $\hat{R}(\mathbf{r})$, supposing a rigid local order.}
\label{circ}
\end{figure}

Following the preceding discussion, we are now equipped to subject an arbitrary insulating material to a spin bias, by one or more voltage-controlled spin reservoirs. This is sketched in Fig.~\ref{circ}, where metallic spin reservoirs are attached to supply arbitrarily oriented spin accumulations $\boldsymbol{\mu}^{(i)}_s$ via, e.g., the spin Hall effect. These spin biases can trigger magnetic dynamics in the material, whose propagation can be detected by one or more output contacts, which operate reciprocally to the input ones.\cite{takeiPRL14,ochoaPRB16} Specifically, we rely here on the Onsager reciprocity,\cite{onsagerPR31p1} according to which, loosely speaking, if a metallic contact can trigger spin dynamics in response to, e.g., an applied current, the same contact should be able to pick up a voltage in response to similar spin dynamics.\cite{volovikJPC87,*barnesPRL07,*duinePRB08sp,*tserkovPRB08mt,*tserkovPRB14}

This philosophy can similarly be employed to study spin currents carried by thermal magnons in magnetic insulators, as has been demonstrated in Refs.~\onlinecite{cornelissenNATP15,*liNATC16}. Here, different platinum contacts were used for injecting and detecting spin flows transmitted by a ferrimagnetic insulator (yttrium iron garnet). According to the bosonic statistics of magnons, this spin-transport regime can be considered to be thermally activated and incoherent. Furthermore, due to a finite lifetime of the spin-carrying excitations, one can generally expect an exponential suppression of the detected signal with distance. In the diffusive transport regime, the latter corresponds to the spin-diffusion length of magnons, $\lambda_s=\sqrt{D\tau_s}$, where $D$ is the diffusion coefficient of thermal magnons and $\tau_s$ is their characteristic lifetime.

\subsection{Spin superfluidity}

More interesting and potentially useful regimes of spin transport concern spin flows that can be carried by coherent order-parameter dynamics, in analogy to charge flows in superconductors, mass flows in superfluid $^4$He, and mass and spin flows in $^3$He.\cite{volovikBOOK03} This can be illustrated by considering an easy-plane magnet, whose local configuration can be parametrized by a canonical pair of variables $(\varphi,\rho_s)$, where $\varphi$ is the polar angle parametrizing the U(1) order-parameter within the easy plane and $\rho_s$ the (nonequilibrium component of the) spin density out of this plane. The canonical conjugacy is evident as $\rho_s$ is the generator of rotations within the easy plane.\cite{halperinPR69} The simplest Hamiltonian describing a smooth order-parameter field is
\eq{
H=\frac{\rho_s^2}{2\chi}+\frac{A(\boldsymbol{\nabla}\varphi)^2}{2}\,,
\label{H}}
where we truncated the expansion at the leading, quadratic order in the deviations from the equilibrium. $A$ here is the order-parameter stiffness against long-wavelength distortions and $\chi$ is the local spin susceptibility. (Supposing the spin-rotational symmetry within the easy plane, the Hamiltonian should not depend on the absolute value of $\varphi$.) The corresponding Hamilton equations of motion are given by
\eq{
\partial_t\varphi\equiv\delta_{\rho_s} H=\frac{\rho_s}{\chi}~~~{\rm and}~~~\partial_t\rho_s\equiv-\delta_\varphi H=A\nabla^2\varphi\,.
\label{dd}}
The first equation can be interpreted as the Josephson relation for the phase $\varphi$, while the second equation can be understood as the continuity equation:
\eq{
\partial_t\rho_s+\boldsymbol{\nabla}\cdot\mathbf{j}_s=0\,,\,\,\,{\rm where}~~~\mathbf{j}_s\equiv-A\boldsymbol{\nabla}\varphi\,.
\label{ce}}
The underlying conservation law is dictated by the symmetry under uniform rotations within the easy plane. The boundary conditions at an interface with a spin-biased metal can be obtained from Eq.~\eqref{Js}, in the case of a collinear local order [or, more generally, from Eq.~\eqref{R}]. Projecting this on the easy-plane dynamics and supposing $\boldsymbol{\mu}_s$ is parallel to the hard axis, we get\cite{takeiPRL14}
\eq{
j_s=g\left(\mu_s-\hbar\partial_t\varphi\right)\,,
}
where the spin conductance $g$ is normalized per unit area of the interface. This is closely analogous to Andreev reflection at a metal/superconductor interface, which is $\propto(2eV-\hbar\partial_t\varphi)$, in terms of the voltage $V$ applied to the normal metal and phase $\varphi$ dynamics of the condensate.

Combining Eqs.~\eqref{dd} results in the wave equation for angular dynamics:
\eq{
\left(\partial^2_t-u^2\nabla^2\right)\varphi=0\,,
\label{wave}}
with the sound velocity $u\equiv\sqrt{A/\chi}$. The linearly-dispersing elementary excitations are akin to the first sound in a neutral superfluid.

\subsection{Role of anisotropies and dissipation}

With the above idealized discussion setting the stage for a superfluid-like treatment of easy-plane spin dynamics, there are at least two ways in which it will differ from the genuine superfluidity, in practice. The crux of the matter is that the latter is rooted in the fundamental gauge symmetry of the underlying condensate, while the former is constructed in terms of an approximate (structural) U(1) symmetry.\cite{kohnRMP70} Breaking this symmetry microscopically, while preserving it on average, introduces a Rayleigh-Gilbert damping\cite{gilbertIEEEM04} $R=\alpha s(\partial_t\varphi)^2/2$ ($\alpha$ being a dimensionless parameter and $s$ a normalization prefactor in units of spin density), which modifies the Hamilton equation for spin density as $\partial_t\rho_s\equiv-\delta_\varphi H-\delta_{\partial_t\varphi}R$. This spoils the continuity equation:
\eq{
\partial_t\rho_s+\boldsymbol{\nabla}\cdot\mathbf{j}_s=-\frac{\rho_s}{\tau_\alpha}\,,
}
where $\tau_\alpha\equiv\chi/\alpha s$ is understood as the spin relaxation time.

\begin{figure*}[!htb]
\includegraphics[width=0.7\linewidth]{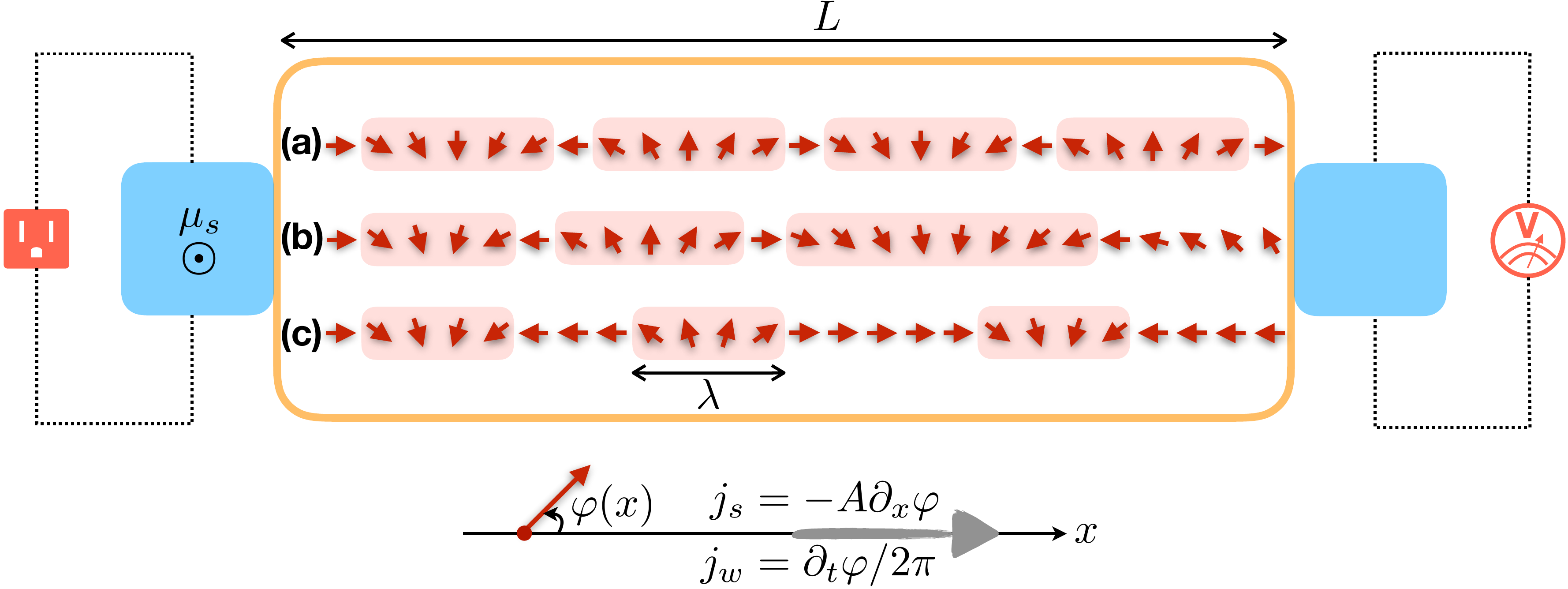}
\caption{Three regimes of collective spin and winding flows, $j_s$ and $j_w$, from the injector terminal (left) to the detector (right), connected by a quasi-one-dimensional easy-plane magnetic strip of length $L$: (a) A perfect spin superflow, where the winding gradient $\partial_x\varphi$ is uniform in a dynamic steady state. (b) Turning on Gilbert damping $\alpha$ introduces a negative gradient in $j_s$, accounting for the leakage of the angular momentum to the substrate. (c) Additionally, adding a small easy-axis anisotropy $K$ along the $x$ axis disrupts a smooth spin flow, by breaking the spin texture down into topological solitons of size $\lambda\sim\sqrt{A/K}$. A steady-state motion of such solitons requires a diminishing pressure as they move along the $x$ axis, corresponding to their decreasing density. The shaded regions highlight magnetic textures with the net (winding) charge of $Q_w=+1/2$. In all three cases, there is a net topological charge flow $j_w$ to the right.}
\label{flow}
\end{figure*}

Breaking, furthermore, the spin-rotational symmetry macroscopically adds anisotropies to the energy \eqref{H}, which can now depend on the absolute value of $\varphi$. For example, introducing an easy-axis anisotropy within the easy plane results in $H(\varphi)\to H(\varphi)-K\cos^2\varphi$. This, together with the above damping, turns the wave equation \eqref{wave} into the damped sine-Gordon equation:\cite{soninJETP78,*soninAP10}
\eq{
\left(\partial^2_t+\tau_\alpha^{-1}\partial_t-u^2\nabla^2\right)\varphi+\frac{K}{\chi}\sin2\varphi=0\,.
\label{sg}}
Injecting a spin current, as before, at an end of such a system will now trigger dynamics that are qualitatively distinct from an ordinary superflow. Rather than simply generating a uniform spiraling flow (in a steady state), there is now a finite threshold for inducing the dynamics (if we neglect, for the moment, thermal activation\cite{kimPRB15br}), upon which a train of (domain-wall) solitons of size $\lambda\sim\sqrt{A/K}$ propagates away from the injector. Their density $n$ grows upon increasing the input bias, coalescing into a state that mimics the original superflow, when $n\sim\lambda^{-1}$.\cite{soninJETP78,konigPRL01} As the pressure needed to push the train (against the viscous Gilbert damping) decreases away from source, the steady-state soliton density will also decrease. Different stages of the collective spin-flow evolution from the perfect superfluid (a), as we turn on the magnetic anisotropies microscopically (b) and macroscopically (c), are illustrated in Fig.~\ref{flow}.

We remark, in the passing, that for the internal consistency of the above discussion, the easy-plane anisotropy needs to be stronger than the parasitic anisotropy $K$. In this case, the aforementioned threshold bias is lower than the upper critical bias dictated by the (Landau) stability of the steady state against small perturbations.\cite{soninJETP78,konigPRL01}

\subsection{Topological-charge hydrodynamics in 1D}

Adding Gilbert damping and macroscopic anisotropies to an idealized spin superflow \eqref{ce} introduces additional terms that spoil the continuity equation for spin dynamics [cf. Eq.~\eqref{sg}]. In one spatial dimension (1D), this results in a viscous solitonic transport, which, at a finite temperature and dilute limit, may be expected to generically exhibit Brownian motion.\cite{kimPRB15br} It turns out, however, that even in this regime, a hydrodynamic description in terms of a robust conservation law is possible. To this end, we are switching from the hydrodynamics of spin density, $\rho_s=\chi\partial_t\varphi$, which is no longer conserved, to the (dual) hydrodynamics of the winding density, $\rho_w\equiv-\partial_x\varphi/2\pi$, which is conserved, as long as the large-angle out-of-plane excursions of the order parameter are penalized by a strong easy-plane anisotropy and can be neglected. Irrespective of the details of the damping and weak in-plane anisotropies, the continuity equation,
\eq{
\partial_t\rho_w+\partial_xj_w=0\,,
\label{cew}}
with $j_w\equiv\partial_t\varphi/2\pi$, is automatically satisfied for a quasi-1D  spin texture, so long as the azimuthal angle $\varphi(x,t)$ is well defined. This is guaranteed if the order parameter never crosses the north or south pole in spin space. The precise conditions for these are dictated by the energetics, the strength of the driving, and thermal fluctuations.

The continuity equation \eqref{cew} sets the departure point for constructing topological hydrodynamics, namely, a transport theory for the conserved topological density, $\rho_w(x,t)$. A natural way to understand the injection mechanism for the associated flow $j_w$ is offered by the energetic considerations. (The detection then follows generally from the Onsager reciprocity.) Namely, projecting the spin-transfer power \eqref{W} onto the easy-plane dynamics, we get
\eq{
\dot{W}=\frac{g}{2\pi}\left[\mu_s\partial_t\varphi-\hbar(\partial_t\varphi)^2\right]=g\left(\mu_sj_w-hj_w^2\right)\,.
}
The first term, $\propto\mu_s j_w$, stems from the torque by the spin bias $\mu_s$ applied to the adjacent reservoir. It is formally analogous to the input power $P=VI$ of an electronic circuit subjected to voltage $V$, when it carries charge current $I$. The second term $\propto-j_w^2$ describes dissipation due to spin pumping,\cite{tserkovRMP05} which is analogous to Joule heating in the electronic counterpart. We thus see that applying a spin bias $\mu_s$ normal to an easy-plane magnet, translates into an energetic bias for the injection of the topological flow $j_w$. This would generate dynamic magnetic textures as those depicted in Fig.~\ref{flow}, with the details governed by magnetic anisotropies and damping. We emphasize that this hydrodynamic construction is dictated entirely by the topology associated with the winding dynamics, not making any simplifying assumptions about the material and structural symmetries of the system.

By the Onsager reciprocity, if the spin bias $\mu_s$ injects flow $j_w$ (e.g., at the left contact depicted in Fig.~\ref{flow}), the topological outflow $j_w$ at the right contact will eject spin current $\propto j_w$,\cite{kimPRB15br} which would in turn generate a measurable voltage $V$ by the inverse spin Hall effect.\cite{hoffmannIEEEM13} The value of $j_w$, in the steady state, is determined by the microscopic details of the magnetic conduit of the topological density $\rho_w$. In a number of generic cases,\cite{takeiPRL14,kimPRB15br} however, it can be written in linear response as
\eq{
j_w\propto\frac{\mu_s}{r_l+r_r+r_b}\,,
\label{jwL}}
where $r_{l,r}$ parametrize the injection impedance at the contacts and $r_b\propto L$ the bulk impedance for the propagation of the winding density along the magnetic channel of length $L$ (cf. Fig.~\ref{flow}). For the idealized spin-superflow regime [Fig.~\ref{flow}(a)],\cite{takeiPRL14} $r= g^{-1}$, at each interface, while $r_b=0$. This mimics an electronic normal-metal/superconductor/normal-metal heterostructure,\cite{nazarovBOOK09} with $g$ replacing the contact Andreev conductance. Adding a Gilbert damping $\alpha$ to this [Fig.~\ref{flow}(b)] gives\cite{takeiPRL14} $r_b\propto\alpha L$, reflecting the leakage of the angular momentum into the substrate at a rate that scales with the system size (since, in the steady state of coherent dynamics, $\partial_t\varphi$ must be uniform throughout the system). Finally, adding in-plane anisotropies [Fig.~\ref{flow}(c)], results in\cite{kimPRB15br} $r_b\propto L/D$, where $D$ is the diffusion coefficient of the domain-wall solitons. Within the Landau-Lifshitz-Gilbert phenomenology of magnetic dynamics,\cite{landauBOOKv9,gilbertIEEEM04} $D\propto\alpha^{-1}$, which can be further modified by pinning effects and the associated creep transport in disordered wires.\cite{lemerlePRL98} In this solitonic case, at elevated temperatures (so quantum-tunneling effects play no role), the proportionality coefficient in Eq.~\eqref{jwL} involves a Boltzmann factor $e^{-\beta E_{\rm dw}}$, where $E_{\rm dw}$ is the free-energy cost to add a single domain wall into a uniform system. The topological flow thus gets exponentially suppressed at low temperatures, as the solitons, which carry both the winding density $\rho_w$ and its flow $j_w$, get depleted from the magnetic wire. As already mentioned,\cite{soninJETP78,konigPRL01} a threshold bias then needs to be applied in order to overcome the energy barrier $E_{\rm dw}$ for injecting domain walls. Above this critical bias, the solitons fill the system and establish a collective drift towards the detector [cf. Fig.~\ref{flow}(c)].

One salient feature of the collective response underlying Eq.~\eqref{jwL} concerns the algebraic, $j_w\propto L^{-1}$, scaling of the nonlocal response, in the limit of $L\to\infty$. This is in stark contrast to the exponential suppression of the signals mediated by a diffusive spin transport carried by magnons\cite{cornelissenNATP15} or other decaying quasiparticles. Here, in essence, in invoking topological arguments for easy-plane dynamics, we have supposed that magnetic solitons (or some arbitrary winding) have an infinite lifetime. In reality, however, this lifetime is effectively finite, albeit exponentially long, $\propto e^{\beta E_\varphi}$, where $E_\varphi$ is the energy barrier for thermally-activated phase slips.\cite{halperinIJMPB10} These correspond microscopically to strong local deviations of the magnetic order away from the easy plane, reaching the north/south poles (in spin space) and thus undoing the winding density $\rho_w$.\cite{kimPRB16} In the limits depicted in Fig.~\ref{flow}(a,b), such phase slips can locally unwind the smooth winding density, while in Fig.~\ref{flow}(c) they can flip the chirality (and thus the sign of the topological charge, $\pm1/2$) associated with each domain wall or spontaneously produce or annihilate pairs of domain walls with the same chirality.

To summarize, the topological protection relies on a large energy barrier $E_\varphi$, which sets an exponentially long lengthscale $e^{\beta E_\varphi}$ for the validity of the continuity equation \eqref{cew} and the associated topological hydrodynamics. We do not expect the nonlocal algebraic signals \eqref{jwL} to persist beyond this lengthscale. It is useful to remark that in the case when the solitonic transport of Fig~\ref{flow}(c) is itself thermally activated,\cite{kimPRB15br} solitonic diffusion that preserves topological charge can be established at intermediate temperatures, $E_{\rm dw}<k_BT<E_\varphi$. The beneficial disparity $E_{\rm dw}\ll E_\varphi$ is generally guaranteed, so long as the dominant magnetic anisotropy in the system is of the easy-plane type (which is naturally assumed throughout). This follows from the dependence $E\propto\sqrt{K}$, for either of these two energies, on the relevant anisotropy $K$.\cite{kimPRB15br,kimPRB16} At very low temperatures, quantum phase slips ultimately take over in relaxing phase winding.\cite{zaikinPRL97} In magnetic systems, this can be sensitive to microscopic details and, in particular, on whether the constituent spins are integer or half-odd-integer.\cite{kimPRL17} Apart from this, the quantum regime of topological hydrodynamics remains largely unexplored. It should be clear, e.g., from the coherent-spin path-integral perspective,\cite{altlandBOOK10} that at least some of the robust features underlying the continuity equation \eqref{cew} and the ensuing long-range transport should survive in the extreme quantum regimes.

\subsection{Higher-dimensional generalizations}

One immediate generalization of the (topological) winding hydrodynamics follows the structure of the homotopy group
\eq{
\pi_n(S^n)=\mathbb{Z}\,.
\label{homo}}
For $n=1$, the integer corresponds to the number of the winding twists discussed in the above one-dimensional case. For $n=2$, this generalizes to the number of skyrmions that characterize topological classes of two-dimensional magnetic textures.\cite{belavinJETPL75} For $n=3$, the underlying topological textures (in three spatial dimensions) are realized by placing the order parameter on a hypersphere.\cite{skyrmeNP62} Alternatively, and more relevant for spin systems, the order-parameter space here may be given by SO(3), i.e., the group of rigid rotations in Euclidean space. This is because $\pi_3(S^3)=\pi_3(\textrm{SO(3)})$, with SO(3) being equivalent (according to the quaternion representation) to the (real) projective space $\mathbb{RP}^3$, so essentially a 3-sphere (with diametrically opposite points identified). One potential physical realization of this is provided by the coherent spin glasses\cite{ochoaPRB18} (or analogous noncollinear frustrated spin systems\cite{dombrePRB89}), in which three independent rotations of random but locally frozen magnetic textures yield three phononic (Goldstone-mode like) branches.\cite{halperinPRB77}

We will illustrate a generalization of the winding hydrodynamics ($n=1$) to higher dimensions, as guided by the homotopy \eqref{homo}, by considering the next simplest case of $n=2$. Physically, this concerns nonlinear $\sigma$ models (such as Heisenberg ferro- or antiferromagnet) in two spatial dimensions. The skyrmionic 3-current $j_{\rm sk}$ underlying the topological hydrodynamics is given by\cite{nakaharaBOOK03}
\eq{
j_{\rm sk}^\mu=\frac{1}{8\pi}\epsilon^{\mu\nu\rho}\epsilon^{abc}l_a\partial_\nu l_b\partial_\rho l_c\,.
\label{jmu}}
Here, $|\mathbf{l}(x,y,t)|\equiv1$ describes a directional order-parameter field. The fully-antisymmetric Levi-Civita symbols $\epsilon$ are accompanied with summations over repeated indices, with the Greek letters $\nu,\mu,\rho$ labeling three space-time coordinates and the Roman letters $a,b,c$ designating three spin-space components. One easily checks that the current \eqref{jmu} obeys the continuity equation:
\eq{
\partial_\mu j_{\rm sk}^\mu=0\,.
}
The conserved (topological) charge,
\eq{
Q_{\rm sk}\equiv\int dxdy\,j_{\rm sk}^0=\frac{1}{4\pi}\int dxdy\,\mathbf{l}\cdot\partial_x\mathbf{l}\times\partial_y\mathbf{l}\,,
\label{Qsk}}
can be recognized as the skyrmion number, which is quantized in integer values (if the order parameter is fixed at the boundary or at infinity to point in the same direction).\cite{belavinJETPL75} This integer is the degree of the $\mathbb{R}^2\to S^2$ mapping, corresponding to the number of times the sphere is covered by the magnetic texture. $Q_{\rm sk}$ can be thought as the two-dimensional generalization of the winding number, which is the degree of the $\mathbb{R}^1\to S^1$ mapping.

In the special case of a ferromagnetic order parameter $\mathbf{l}$, we can easily establish an energetic bias for the skyrmionic spin injection from a metallic contact using the adiabatic spin-transfer torque.\cite{tataraPRP08} Namely, applying an electric current $\vec{j}$ tangential to the interface, the torque (per unit length of the contact)
\eq{
\boldsymbol{\tau}=\frac{\hbar\mathcal{P}}{2e}\vec{j}\cdot\vec{\nabla}\mathbf{l}
\label{stt}}
would generally arise in the proximity to a smooth magnetic texture. This torque follows from the (proximal) exchange interaction between electrons in the normal-metal contact and the (insulating) ferromagnet. $\mathcal{P}$ is a dimensionless parameter parametrizing the strength of this exchange (with $|\mathcal{P}|\to1$ in the extreme case of a very strong interaction that would polarize and lock electron spins to the magnetic texture\cite{tataraPRP08}).

The work done by the torque \eqref{stt} can be evaluated to yield the power
\eq{
\dot{W}=\int dr\,\boldsymbol{\tau}\cdot\mathbf{l}\times\dot{\mathbf{l}}=\frac{h\mathcal{P}j}{e}\hat{z}\cdot\int d\vec{r}\times\vec{j}_{\rm sk}\,,
\label{Wsk}}
where the integration is performed along the length of the current-$j$ carrying contact. We see from this that the electric current tangential to a magnetic interface produces an energetic bias for the transverse skyrmion-density injection. We can thus expect that a nonequilibrium skyrmion charge \eqref{Qsk} would generally develop over time, in the presence of such a bias. The details of the efficiency of this skyrmion injection depend of course on the physical regime of the system. In particular, such skyrmionic injection and subsequent flow were studied in Ref.~\onlinecite{ochoaPRB16} in the regime of a thermally-activated Brownian motion of a dilute gas of rigid (solitonic) skyrmions. In Ref.~\onlinecite{ochoaPRB17}, the ensuing skyrmion flow was suggested as a probe for different textured phases of chiral magnets (such as collinear, helical, and skyrmion-crystal phases), which would yield different skyrmionic responses. In particular, in the crystalline phase, the work \eqref{Wsk} would translate into a boundary pressure that could trigger a gyrotropic sliding motion of the skyrmionic crystal as a whole. One could easily envision other physical scenarios, where such topological hydrodynamic probes may give useful information about a nontrivial magnetic ordering, which would otherwise not be directly accessible via other transport measurements.

\begin{figure}[!htb]
\includegraphics[width=\linewidth]{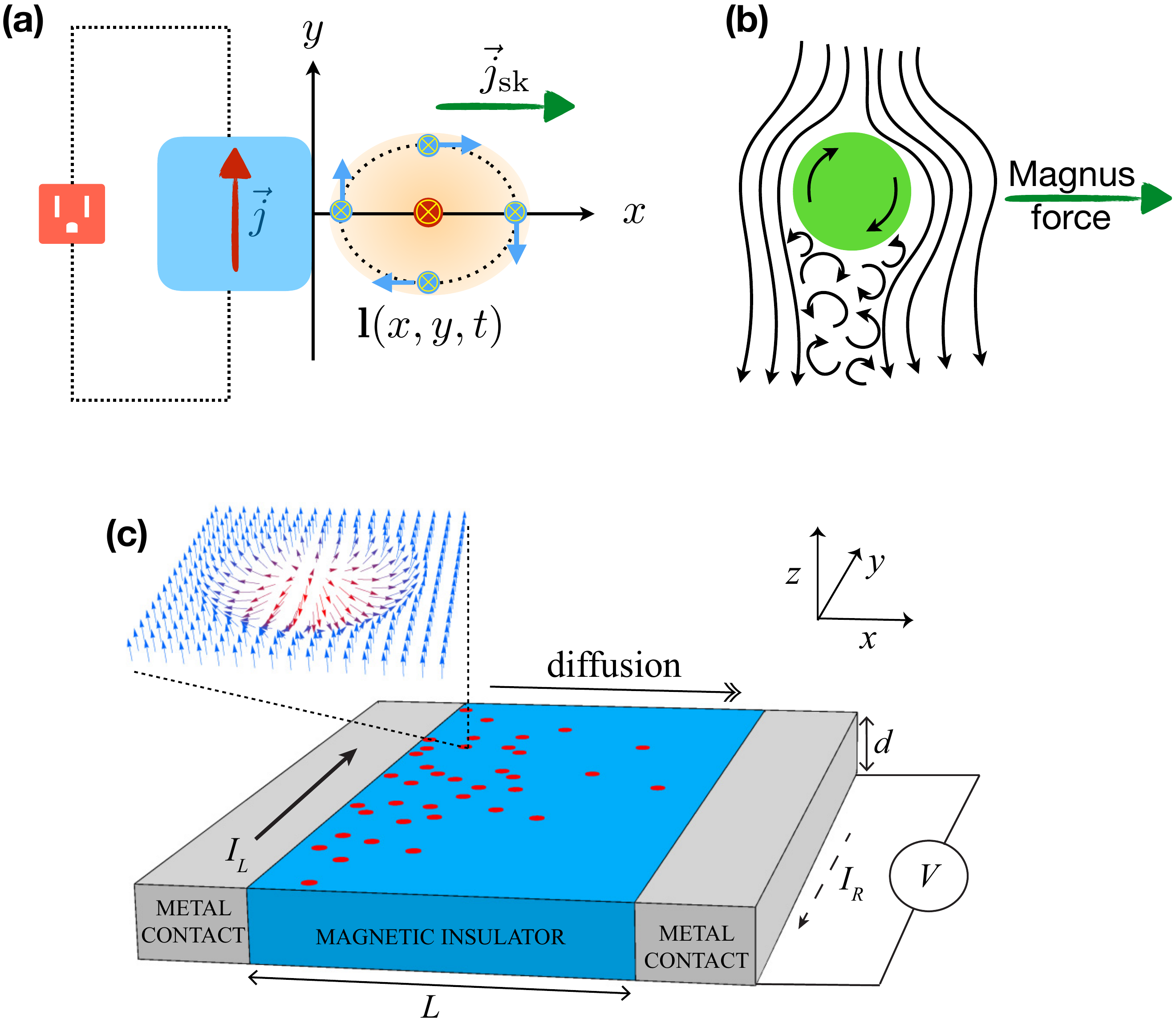}
\caption{(a) Electric-current induced injection of skyrmion flux into the magnetic region ($x>0$), according to the work \eqref{Wsk}. Geometrically, this is analogous to the Magnus force (b). Panel (c) shows a four-terminal electrical (drag) transconductance measurement, which could detect the injected (nonequilibrium) skyrmion flow from the left to the right metal.\cite{ochoaPRB16}}
\label{magnus}
\end{figure}

In Fig.~\ref{magnus}(a), we schematically depict this spin-torque-induced skyrmion injection into a magnetic insulator. The latter could be either an ordinary Heisenberg ferromagnet or a chiral magnet with propensity to form skyrmion textures due to the Dzyaloshinski-Moriya interaction.\cite{bogdanovJMMM94} Panel (b) of the figure illustrates geometrical analogy between the current-induced skyrmion flow and the Magnus force (which is produced by the turbulent wake aft of a rotating body subjected to a hydrodynamic flow). Panel (c) (cf. Ref.~\onlinecite{ochoaPRB16} for more details) shows a nonlocal electrical measurement, which probes a nonequilibrium skyrmion flux between two metal contacts. Similarly to Fig.~\ref{flow}, the left metal contact injects the topological hydrodynamics (now of the skyrmionic flavor). The right contact detects an electromotive force $\mathcal{E}$ produced by the skyrmionic outflow through the right contact, as dictated by the Onsager reciprocity:\cite{volovikJPC87}
\eq{
\mathcal{E}=\frac{\hbar\mathcal{P}}{2e}\int d\vec{r}\,\mathbf{l}\cdot\vec{\nabla}\mathbf{l}\times\dot{\mathbf{l}}=-\frac{h\mathcal{P}}{e}\hat{z}\cdot\int d\vec{r}\times\vec{j}_{\rm sk}\,.
}
In the diffusive regime of solitonic propagation of skyrmion density, as sketched in Fig.~\ref{magnus}(c), the resultant transconductance scales algebraically as $L^{-1}$ with the length $L$ of the topological transport channel, similarly to the previous winding example, Eq.~\eqref{jwL}. This stems from the conserved character of the topological flow and the generic (Ohmic) scaling $\propto L$ of its impedance. The latter is determined by the solitonic diffusion coefficient, which depends on Gilbert damping, impurity potential, etc.

\section{Summary and discussion}

Spin transport in magnetic insulators may be carried either by spin-carrying quasiparticles, such as magnons in ordered spin systems, or coherent order-parameter dynamics, such as an easy-plane superflow. In either case, the notion is strictly-speaking meaningful when there is a spin-rotation symmetry axis, along which the spin angular momentum is conserved, at least approximately. It is remarkable that, while the continuity equation for spin flow breaks down in the opposite regime, broad classes of magnetic materials may still exhibit robust collective transport behavior. The latter can emerge, for example, when the real-space order-parameter textures can be classified into classes distinguished by an extensive topological invariant. Here, we illustrated this by focusing on two simple examples: winding dynamics in one spatial dimension and skyrmion dynamics in two dimensions. Noncollinear magnetic textures parametrized by three Euler angles can allow to also extend these ideas to three-dimensional materials, such as spin glasses.\cite{tserkovPRB17,ochoaPRB18}

One could envision also other types of topological hydrodynamics, which could be guided by the homotopy considerations for the coherent order-parameter fields. With the key relevant mathematical concepts already established in other areas of research, including both high and low energies,\cite{volovikBOOK03} the tools of spintronics are opening opportunities to explore broad classes of magnetic materials from the perspective of topological transport. The first steps in this direction are already being made.\cite{wesenbergNATP17,yuanSA18,stepanovNATP18} The topological hydrodynamics appears appealing both as a tool to probe complex phases of quantum materials\cite{stepanovNATP18} and, eventually, as a utilitarian resource within spintronics.\cite{tserkovPRL18}

\begin{acknowledgments}
I am grateful to Benedetta Flebus, Se Kwon Kim, Hector Ochoa, So Takei, Pramey Upadhyaya, and Ricardo Zarzuela for insightful discussions and collaborations. The work was supported in part by the NSF under Grant No. DMR-1742928 and the ARO under Contract No. W911NF-14-1-0016.
\end{acknowledgments}

\end{document}